\documentclass[12pt]{iopart}

\usepackage{amsgen}
\usepackage{amssymb}
\usepackage{graphicx}
\usepackage{setstack}
\usepackage{color}

\def\be{\begin{equation}}
\def\ee{\end{equation}}
\def\beq{\begin{equation}}
\def\eeq{\end{equation}}
\def\bea{\begin{eqnarray}}
\def\eea{\end{eqnarray}}

\makeatletter
\@ifundefined{textcolor}{}
{
 \definecolor{BLACK}{gray}{0}
 \definecolor{WHITE}{gray}{1}
 \definecolor{RED}{rgb}{1,0,0}
 \definecolor{GREEN}{rgb}{0,1,0}
 \definecolor{BLUE}{rgb}{0,0,1}
 \definecolor{CYAN}{cmyk}{1,0,0,0}
 \definecolor{MAGENTA}{cmyk}{0,1,0,0}
 \definecolor{YELLOW}{cmyk}{0,0,1,0}
 }

\newcommand{\negminispace}{\kern-.016667em} % minispaces % = 1/10 of \negthinspace % In amsmath, use \mspace{1mu} = 1/3 \thinspace = 1/18 em

\newcommand{\half}{\kern.083333em}   % 1/2 of \thinspace
\newcommand{\quart}{\kern.0416675em}  % 1/4 of \thinspace
\newcommand{\nhalf}{\kern-.083333em}   % 1/2 of \negthinspace
\newcommand{\nquart}{\kern-.0416675em}  % 1/4 of \negthinspace

\begin{document}

\title{The Method of Images in Cosmology}

\author{Timothy Clifton}
\address{School of Physics and Astronomy, Queen Mary University of London, UK.
}

%\eads{\mailto{}}

%\pacs{98.80.Jk, 04.20.Cv}

\begin{abstract}

We apply the method of images to the exact initial data for cosmological models that contain a number of regularly arranged discrete masses. This allows us to join cosmological regions together by throats, and to construct wormholes in the initial data. These wormholes allow for the removal of the asymptotically flat ``flange'' regions that would otherwise exist on the far side of black holes. The method of images also provides us with a way to investigate the definition of mass is cosmology, and the cosmological consequences of the gravitational interaction energies between massive objects. We find evidence that the interaction energies within clusters of massive objects do indeed appear to contribute to the total energy budget in the cosmological regions of the space-time.

\end{abstract}

\section{Introduction}
\label{sec:int}

A recent development in the field of relativistic cosmology is a renewed interest in the study of models that contain discrete masses. These models do not assume a matter content that takes the form of a fluid, but instead take the mass in the universe to be contained within a number of (usually) regularly distributed black holes. The primary motivation for these studies has often been stated as the desire to remove the need for averaging from cosmological modelling, and hence to allow for the exploration of ideas about the influence of structure formation on the large-scale evolution of the Universe \cite{backreaction,backreaction2}. They also, however, allow one to consider problems such as the behaviour of black holes in an evolving universe, and the definition of mass in cosmology.

One of the first papers on this subject was the pioneering work of Lindquist and Wheeler, who discussed how time-symmetric initial data could be constructed for just such a universe, as well as providing an aproximate framework for calculating its evolution \cite{LW}. These ideas were recently revived in \cite{CF1,CF2}, where the optical properties of the approximate models were studied in detail. This was followed by an in depth investigation of the exact initial data, including a precise numerical calculations of the scale of such solutions \cite{CRT12}. For small numbers of black holes ($\lesssim 10$) it was found that deviations on the order $\sim 10\%$ exist from comparable dust-dominated Friedmann models with the same total mass. This difference, however, was shown to decrease to $\lesssim 1 \%$ as the number of masses becomes large ($\gtrsim 100$). In suitable limits, it has also now been shown that the scale of these discrete models converges on that of their Friedmann counterparts as the number of masses in the space-time diverges \cite{korzy}.

Further recent studies in this area have considered the numerical evolution of a universe with $8$ regularly spaced black holes on a topological 3-sphere \cite{BK}, the exact evolution of curves that are equidistant from all nearby masses \cite{CRTG13}, and the behaviour of the geometry of space-time near surfaces that exhibit a reflection symmetry \cite{CRG14}. Numerical studies have also been performed for the case of infinitely many masses arranged on a cubic lattice \cite{cube1, cube2, cube3, cube4, cube5}. This work has all contributed to a deeper understanding of the problem of the relativistic modelling of a Universe with matter that is clumped into largely isolated masses, but there remain a number of issues that are still not properly understood.

In this paper we use the method of images to construct new sets of time-symmetric initial data. Much of this work proceeds along the same lines as that of Misner, who studied the application of the method of images to asymptotically flat time-symmetric initial data \cite{Misner63}. By developing these methods for cosmological initia data we are able to create multiple cosmological regions that are images of each other, and that are connected by throats. This allows us to consider the cosmological consequences of the interaction energies between black holes, and can be used to construct initial data that contains wormholes ({\it i.e.} is non-simply connected). These wormholes remove the need to have causally disconnect regions when considering time-symmetric initial data for cosmological models, and can result in compact vacuum universes.

In Section 2 we recap the use of geometrostatics to construct time-symmetric initial data in cosmology. In Section 3 we discuss the definitions of mass and energy in these models. In this section we also introduce a two scale problem which we believe to clarify at least some of the cosmological consequences of interaction energies in cosmology. Section 4 contains the derivation of reflection operators and maps that can be used to implement the method of images in cosmology. We then proceed to explicitly calculate the geometries that result from applying this method to a cosmological model containing $8$ black holes regularly arranged on a 3-sphere. These geometries allow us to investigate the effects of interaction energies in explicit geometries, and to construct wormholes that connect antipodal black holes. In Section 6 we then construct a compact universe in which all antipodal black holes are connected by wormholes. Finally, we summarize with a discussion of our results in Section 7.

\section{Time-Symmetric Cosmological Solutions}

In general relativity the geometry of a 3-dimensional space cannot be specified arbitrarily, but must be chosen to satisfy the constraint equations, $G_{i}^{\phantom{i}0}= 8 \pi G T_{i}^{\phantom{i}0}$. In vacuum these equations can be written
\bea
\label{constraint1}
\mathcal{R} +K^2 - K_{ij} K^{ij} = 0\\
(K_i^{\phantom{i}j} - \delta_i^{\phantom{i} j} K )_{\vert j} = 0,
\label{constraint2}
\eea
where $i,j$ run over spatial indices, $\mathcal{R}$ is the scalar curvature of the spatial metric $g_{ij}$, and $K_{ij}$ is the extrinsic curvature of this 3-dimensional hypersurface in the full 4-dimensional space-time. For an appropriate choice of time coordinate this extrinsic curvature can be written as $K_{ij} = -\frac{1}{2} \partial g_{ij} / \partial t$.

Eqs. (\ref{constraint1}) and (\ref{constraint2}) are in general extremely complicated to solve. The situation is dramatically simplified, however, if we restrict ourselves to considered 3-dimensional geometries that are instantaneously static. In this case the geometry of the 4-dimensional space-time is time-reversal symmetric around the 3-dimensional hypersurface, and we say that the surface itself is ``time-symmetric''. This symmetry leads to the vanishing of the hypersurface's extrinsic curvature, such that $K_{ij}=0$. It can then be seen that  Eqs. (\ref{constraint1}) and (\ref{constraint2}) are satisfied if and only if
\be
\mathcal{R}=0.
\label{constraint1b}
\ee
Any 3-geometry that satisfies this equation is a solution to the vacuum Einstein equations on a time-symmetric hypersurface, and can be used as initial data for the evolution that will constitute the full space-time.

It can be shown that Eq. (\ref{constraint1b}) is satisfied by the following 3-geometry \cite{CRT12}:
\be
\label{geom}
dl^2 = \psi^4 d\sigma^2  \; ,
\ee
where $d\sigma^2$ is the line-element of a unit 3-sphere with metric $h_{ij}$, such that
\be
d\sigma^2 = h_{ij} dx^i dx^j = d \chi^2 +\sin^2 \chi d \theta^2 + \sin^2 \chi \sin^2 \theta d \phi^2 \; .
\ee
In what follows, we will refer to this 3-sphere as our ``reference hypersphere''. This sphere will be useful for marking the positions of points in our 3-dimensional geometry, although the reader should remember that it is only conformally related to the geometry of space through Eq. (\ref{geom}).  In order for the geometry in Eq. (\ref{geom}) to be a solutions of Eq. (\ref{constraint1b}), the conformal factor must satisfy the following equation:
\be
\label{helmholtz}
\nabla^2 \psi \equiv \frac{1}{\sqrt{ \vert h \vert }} \partial_i \left( \sqrt{ \vert h \vert } h^{ij} \partial_j \psi \right) = \frac{3}{4} \psi.
\ee
This is the Helmholtz equation, and is linear in $\psi$. It is remarkable that a linear equation of this type can be found, and it should be noted that this is not due to any approximation or linearisation that we have performed. It is simply that the full constraint equations happen to be linear when time-symmetry is imposed \cite{Misner63}. This property means that solutions to Eq. (\ref{helmholtz}) can be superposed, and the result will also be a solution to Eq. (\ref{constraint1b}).

\subsection{The Schwarzschild Solution}

\begin{figure}[t]
\begin{centering}
\includegraphics[width=15cm]{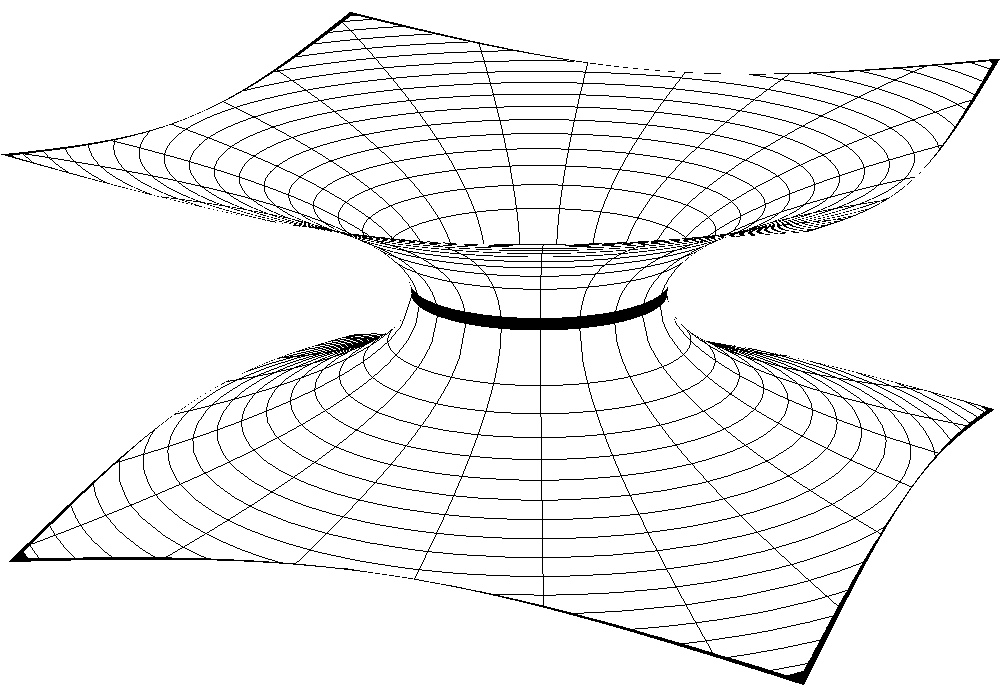}
\par\end{centering}
\caption{An embedding diagram for the $\theta=\pi/2$ section of the time-symmetric initial data of the Schwarzschild solution, as given by Eqs. (\ref{geom}) and (\ref{schw}). The poles at $\chi=0$ and $\chi=\pi$ occur at asymptotic distances in the upper and lower parts of the plot. The thick band corresponds to a slice through the MOTS at $\chi=\pi/2$.}
\centering{}\label{zerospheres}
\end{figure}

The Schwarzschild solution to Einstein's equations admits a time-symmetric initial value problem, with an intrinsic geometry given by Eq. (\ref{geom}) with
\be
\label{schw}
\psi = \frac{\sqrt{m}}{2 \sin \frac{\chi}{2}} + \frac{\sqrt{m}}{2 \cos \frac{\chi}{2}},
\ee
where $m=$constant is the usual Schwarzschild mass parameter. This form of the conformal factor can be seen to satisfy Eq. (\ref{helmholtz}), and appears to suggest that the time-symmetric Schwarzschild solution can be considered as two masses positioned at antipodes on our reference hypersphere (one at $\chi=0$, and the other at $\chi=\pi$). The position of the Marginally Outer Trapped Surfaces (MOTS) corresponding to these two masses are coincident, and occur at $\chi = \pi/2$. One can then consider the region $\chi<0$ to be the region exterior to the horizon of the mass at $\chi=\pi$, or the region interior to the horizon of the mass at $\chi=0$. Likewise for the region $\chi>\pi/2$, {\it mutatis mutandis}. The embedding diagram for the surface $\theta=\pi/2$ is shown in Fig. \ref{zerospheres}, and is asymptotically flat in the limits $\chi \rightarrow 0$ and $\chi \rightarrow \pi$.

%%%%%%%%%%%%%%%%%%%%%%%%%%%%%
%%%%%%%%%%%%%%%%%%%%%%%%%%%%

\subsection{A Solution With N-Bodies}

As described in \cite{CRT12}, it is possible arrange arbitrarily many point-like masses on our reference hypersphere, at arbitrary locations, by adding extra terms of the form that appear in Eq. (\ref{schw}). The conformal factor for this situation is given by
\be
\label{nbody}
\psi = \sum_{i=1}^N \frac{\sqrt{\tilde{m}_i}}{2 f_i (\chi , \theta, \phi)},
\ee
where $\tilde{m}_i$ are a set of constant mass parameters, and where $f_i(\chi, \theta, \phi) = \sin (\chi_i/2)$. The $\chi_i$ in this expression refers to a set of new coordinates ($\chi_i$, $\theta_i$, $\phi_i$) that are obtained by rotating ($\chi$, $\theta$, $\phi$) until the $i$'th term in Eq. (\ref{nbody}) appears at $\chi_i=0$.

\begin{figure}[t]
\begin{centering}
\includegraphics[width=10cm]{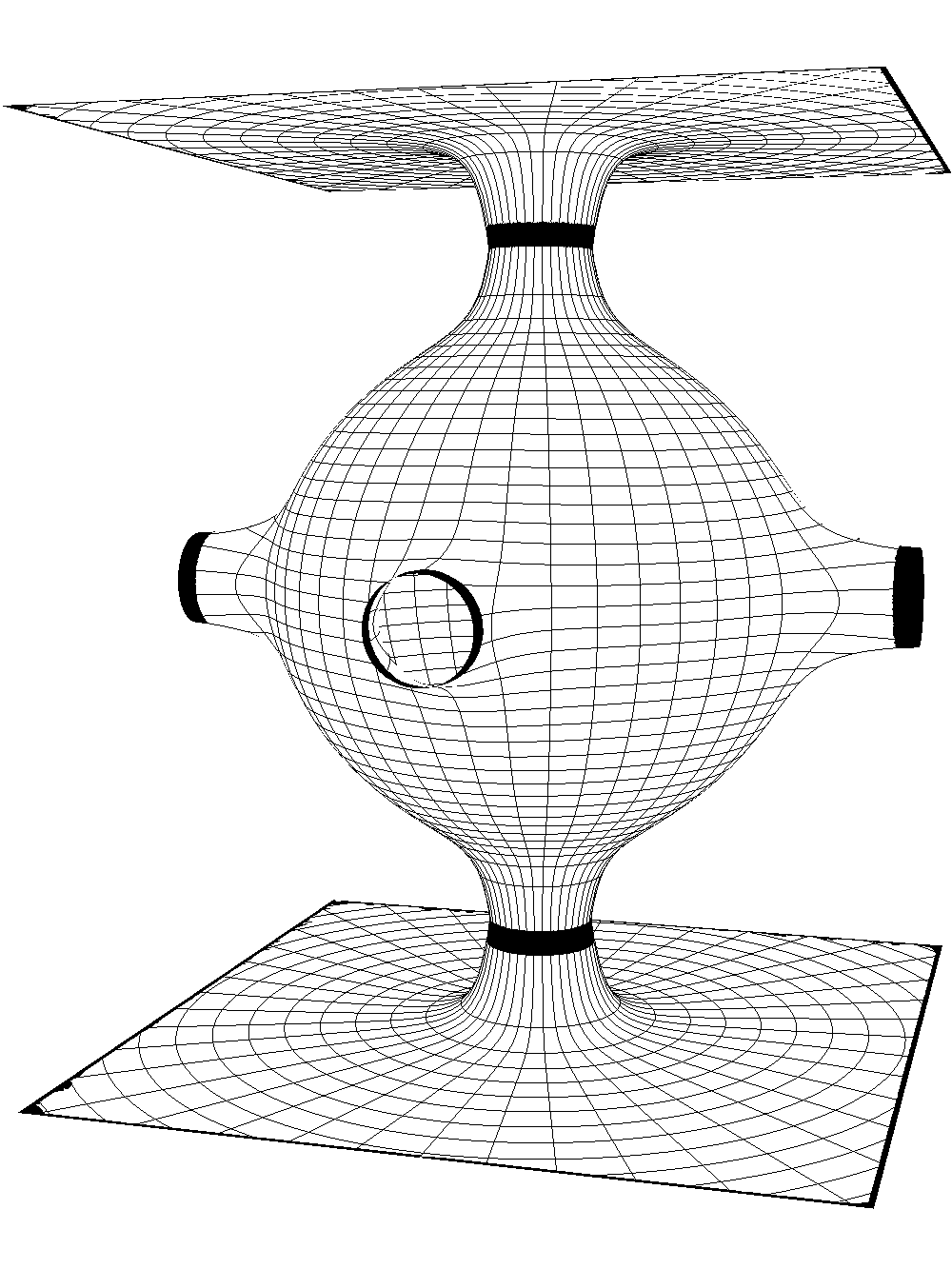}
\par\end{centering}
\caption{An illustration of the embedding diagram for a two-dimensional slice through the initial data of a cosmology, containing $6$ black holes. The central sphere is the cosmological region that contains the black holes, and the asymptotically flat regions are the flanges that occur on the other side of the black hole's MOTS (again shown as thick black bands). Only $2$ of the flange regions are displayed in this illustration, for ease of presentation.}
\centering{}\label{onesphere}
\end{figure}

Although less obvious than in the Schwarzschild case, the geometry described by Eqs. (\ref{geom}) and (\ref{nbody}) is non-singular everywhere. In the limit that $\chi_i \rightarrow 0$ the geometry of space approaches that of the Schwarzschild geometry at the location of the mass point, and is therefore asymptotically flat. The MOTS of each of these masses corresponds to the topological 2-sphere of smallest possible area that encompasses the mass \cite{gibbons72}. If the mass points are positioned far enough apart, and subject to suitable conditions, the region exterior to the set of all MOTS of all masses then approaches the geometry of a 3-sphere as the number of masses is increased \cite{korzy}. We therefore have a geometry of the type that is illustrated in Fig. \ref{onesphere}. Each mass point corresponds to a bridge that connects the cosmological space filled with N-black holes to an asymptotically flat ``flange'' region. The MOTS of each black hole is the minimal surface that can be found along each bridge.

This is a somewhat peculiar situation, as for each black hole that we wish to include in our cosmological model we are forced to include a causally disconnected asymptotically flat region. As the evolution of the space proceeds each of the bridges is expected to collapse \cite{CRTG13}, and $N$ separate asymptotically flat black hole spaces are expected to result, along with our single cosmological space containing $N$ black holes. This is a very wasteful way to construct a cosmological model, and means, for example, that volume averages cannot be constructed for the initial data, as the volume of space is formally divergent. We will find below that the method of images can be applied to this situation to remove the asymptotically flat flanges, and to leave a single cosmological region with finite volume.

%%%%%%%%%%%%%%%%%%%%%%%%%%%%%
%%%%%%%%%%%%%%%%%%%%%%%%%%%%

\section{Interpretation of Mass and Energy}
\label{sec:energy}

Interaction energy in geometrostatics has been well studied, starting with the work of Brill and Lindquist \cite{BL63}. The focus of these previous studies has often been to calculate the gravitational consequences of the interaction energy between multiple black holes in an asymptotically flat space. In such circumstances the total interaction energy between all black holes can be shown to be equal to the difference between the mass inferred at infinity for the entire system of all black holes, and the sum total of the individual masses that would be inferred in each of their respective flange regions, in some suitable limits. That is, one finds that the interaction energy between black holes is itself a source for the gravitational field. Hence, gravity gravitates.

The consequences of interaction energies in cosmology (i.e. in spaces that are not asymptotically flat) has not yet received much attention. In fact, the dust approximation that is commonly used in cosmology explicitly neglects all interactions, and so implicitly neglects any cosmological consequences of the gravitational effects of interaction energies before, during, and after the formation of structure. Some attempt was made in \cite{CRT12} to proceed by analogy to the asymptotically flat case, and led to surprisingly large effects. Whether or not this analogy is justifiable remains to be seen, but it certainly motivates the further study of the possible consequences of interaction energies in cosmology. The present study offers some scope to shed light on this subject, and so we will elaborate on it further here.

\subsection{Effective Mass and Proper Mass}

To proceed with this discussion we first need to define what is meant by ``mass''. As in \cite{CRT12}, we make the following two definitions:

\begin{enumerate}
\item The {\it effective mass} of a gravitational source is defined to be equal to the mass parameter $\tilde{m}_i$ in Eq. (\ref{nbody}).
\item The {\it proper mass} of a gravitational source is defined as being equal to the mass of the Schwarzschild solution that is approached as $\chi_i \rightarrow 0$. It will be denoted by $m_i$.
\end{enumerate}
We will also define:

\begin{enumerate}
\item[(iii)] The {\it charge}\footnote{No connection to electromagnetism is implied by this name. Our masses are all taken to have no electromagnetic charge.} of a pole in $\psi$ is said to be equal to $\sqrt{\tilde{m}_i}$, as this is the quantity that appears in the numerator of each term in Eq. (\ref{nbody}).
\end{enumerate}

In the asymptotically flat case there is no distinction between the ``effective mass'' and the ``charge'' of the poles in the conformal factor. In that case the total effective mass of a system of $N$-black holes (or, equivalently, the total charge) can be shown to be equal to the total proper mass of each of the individual black holes, and the sum of all of the interaction energies between them (for a suitable definition of interaction energy). In the present case, the effective mass of a gravitational source is equal to the square of its charge. In what follows we will investigate the connection between these quantities and the proper masses and interaction energies of the other black holes present in a cosmological region.

The ``proper mass'' of each of the black holes can be thought of as the mass that would be inferred by an observer who is asymptotically far away in one of the flange regions. In \cite{CRT12} and \cite{korzy} it was shown that it is the sum total of the proper masses of all  black holes in the cosmological region that best approximates the mass of a dust-dominated Friedmann-Lema\^{i}tre-Robertson-Walker (FLRW) solution (when they are regularly arranged, at least). This is not too suprising, as the dust approximation itself neglects all interaction energies, as does taking the sum of all proper masses. 

This situation is somewhat troubling for two reasons. Firstly, we want to build a cosmological model that represents the real universe (which contains interactions), rather than a dust-filled Friedmann universe (which does not). The statement that the sum of proper masses approaches the mass of a dust filled Friedmann universe does not necessarily help with this problem. Secondly, the mass of each black hole is being calculated using the geometry of regions of space that no observer in the cosmological region will ever have access to. This makes it impossible for an actual observer in the cosmological region to operationally determine the proper mass of any of the black holes around them. The study that follows will shed some light on both of these problems.

\subsection{Interaction Energy}

Let us first find an explicit expression for the proper mass of a black hole. To do this, we note that in the limit $\chi_i \rightarrow 0$ the conformal factor from Eq. (\ref{geom}) takes the following form:
\be
\psi \rightarrow A_i + \frac{\sqrt{\tilde{m}_i}}{\chi_i}\, , 
\hspace{2cm} {\rm where} \hspace{1cm} 
A_i = \sum_{j \neq i} \frac{\sqrt{\tilde{m}_j}}{2 \sin \left( \frac{\chi_{ij}}{2} \right)} \, ,
\ee
and where $\chi_{ij}$ is the coordinate distance in $\chi_i$ between the mass points labelled by $i$ and $j$. If we now define a new coordinate $\hat{\chi}_i \equiv A_i^{2} \chi_i$, then the line-element for the geometry of space around the point $\chi_i =0$ can be written
\be
ds^2 \rightarrow \left( 1+ \frac{A_i \sqrt{\tilde{m}}_i}{\hat{\chi}_i} \right)^4 \left( d \hat{\chi}_i^2 + \hat{\chi}_i^2 d \Omega_i^2 \right) \, .
\ee
This can be compared to the $r \rightarrow 0$ limit of the Schwarzschild solution,
\be
ds^2 \rightarrow \left( 1+ \frac{m}{2 r} \right)^4 \left( d r^2 + r^2 d \Omega_i^2 \right) \, ,
\ee
to find
\be
\label{properm}
m_i = 2 A_i \sqrt{\tilde{m}_i} = \sum_{j \neq i} \frac{\sqrt{\tilde{m}_i \tilde{m}_j}}{\sin \left( \frac{\chi_{ij}}{2} \right)} \, .
\ee
This expression shows that the proper mass of each black hole is a sum of $N-1$ terms, one for each of the other black holes present in the solution. Each of these terms is proportional to charge of the black hole in question, and each of them is also proportional to the charge of one of the other black holes. Changing the charge on any one pole therefore changes the proper mass of every black hole in the entire universe.

In order to write down an expression for interaction energies, let us consider a situation in which we have a small cluster of $n$ masses, such that the angular separation of any two masses within that cluster is given by $\chi_{ij} \ll 1$, and such that all other masses in the universe are far away. Let us now consider each of these masses individually, as if we were going to place each of them down within the background space that results from the other $N-n$ black holes that exist in the universe. In this case, the proper mass of each of these objects would be given by
\be
\label{mcluster}
\mathfrak{m}_i = 2 B_{i} \sqrt{\tilde{m}_i} \; , \hspace{2cm} {\rm where} \hspace{1cm} B_i \equiv \sum_{j=1}^{N-n} \frac{\sqrt{\tilde{m}_j}}{2 \sin \left(\frac{\chi_{ij}}{2}\right)} \, ,
\ee
where the index $i$ now labels one of the masses in the cluster, and the index $j$ has been taken to run over all masses outside of the cluster. We can similarly calculate what the proper distance would be between any two masses within the cluster, in the geometry of the background space that results from all masses outside of the cluster. For $\chi_{ij} \ll 1$, this will be given by
\be
\label{rcluster}
r_{ij} \simeq B_i^2 \chi_{ij} \; ,
\ee
where we have assumed that the background space is close to flat throughout the spatial extent of the cluster, such that $\psi \simeq B_i$ for any $i$. This approximation should be expected to valid as long as the masses within the cluster are close together, compared to their distance to all mass outside the cluster. A reasonable expression for the sum of all interaction energies within the cluster is then given by
\be
\label{mint}
m_{\rm int} \equiv - \sum_i^n \sum_{j <i} \frac{\mathfrak{m}_i \mathfrak{m}_j}{r_{ij}} \simeq - \sum_i^n \sum_{j\neq i}^{n-1} \frac{\sqrt{\tilde{m}_i \tilde{m}_j}}{\sin \left(  \frac{\chi_{ij}}{2} \right)}   \, ,
\ee
where in deriving the last expression we have used Eqs. (\ref{mcluster}) and (\ref{rcluster}), and the approximations already specified above. Finally, if we add this expression to the total proper mass in the cluster, $m_T = \sum_i^n m_i$, then we get
\be
\hspace{-2cm}M = m_T + m_{\rm int}  
\simeq  \sum_i^n \sum_{j \neq i}^{N-1} \frac{\sqrt{\tilde{m}_i \tilde{m}_j}}{\sin \left( \frac{\chi_{ij}}{2} \right)} - \sum_i^n \sum_{j\neq i}^{n-1} \frac{\sqrt{\tilde{m}_i \tilde{m}_j}}{\sin \left(  \frac{\chi_{ij}}{2} \right)} 
= \sum_i^n \sum_j^{N-n} \frac{\sqrt{ \tilde{m}_i \tilde{m}_j}}{\sin \left(  \frac{\chi_{ij}}{2}\right)} \, .
\label{M}
\ee
In the last expression, the sum over $i$ indicates the sum over all masses within the cluster, while the sum over $j$ indicates sum over all masses outside the cluster. Thus, if we expect both the proper masses and the interaction energies within our cluster to gravitate, then $M$ should be a reasonable approximation to the cluster's total gravitational mass. This is interesting as it gives
\be
\label{M2}
M = \sum_i^n \mathfrak{m}_i = \sum_i^n B_i \sqrt{{\tilde m}_i} \simeq B \sum_i^n \sqrt{{\tilde m}_i} \, ,
\ee
where $B_i \simeq B$ for every mass in the cluster. This shows that the total gravitational mass of the cluster is proportional to the sum of the charges within it. That is, the sum of the charges of a cluster of objects (multipled by $B$) appears to encode information about both the proper mass and the interaction energies within that cluster, just as in the asymptotically flat case \cite{BL63}.

There are a couple of peculiarities in this discussion that we should expand upon here. Firstly, the mass $M$ is not defined in the same way as in asymptotically flat solutions. In those cases it is possible to expand the gravitational field of a cluster of points in powers of $1/r$, and to compare the results of this with the Schwarzschild solution in the same limit. Here, however, there is no asymptotic region, so $M$ does not have such a clear interpretation. A reasonable course of action is therefore to consider if the cosmological region that contains the cluster behaves as if it contains an object of mass $M$, or whether the cosmology is better described by assigning some other notion of mass to the cluster. We will investigate this in what follows. 

The second peculiarity is that the $\mathfrak{m}_i$ used in Eq. (\ref{mint}), and given explicitly in Eq. (\ref{mcluster}), is calculated by summing over the masses outside of the cluster only. Likewise, the proper distance $r_{ij}$ is calculated with reference to the masses outside of the cluster only. If all of the masses in the cluster are separated by many Schwarzschild radii then these should be good approximations to the actual proper mass, and  the proper distance between masses, respectively. Otherwise, one is left in the position of calculating the interaction energy using quantities derived using the geometry of space that would exist if the gravitational effects of the cluster were themselves neglected. This is also true of the corresponding analysis in the asymptotically flat case \cite{BL63}.

The reader may note that the interaction energies considered here are those between objects that are clustered into some small region of space only. The interaction energies between masses separated by large distances, and the interaction energies between separate clusters, have not been dealt with at all. The effects of these quantities on cosmological behaviour is still an open problem. We will return to it in future studies.

\section{Reflection Operators}

In this section we will introduce the idea of reflection maps and operators that act on the positions of masses, and the geometry of space that results. Such operators are known to exist in the asymptotically flat case \cite{Misner63}, in which case the constraint equations reduce to Laplace's equation, and known results from electromagnetism can be applied. They have also begun to be studied in the cosmological situations, where reflections in great 2-spheres are possible \cite{Kjell}. Here we will be interested in reflections in arbitrary 2-spheres, in the cosmological models that consist of masses regularly arranged on a reference hypersphere. We therefore need to derive new expressions.

\subsection{The Schwarzschild Geometry}

The well known isometry that exists in the time-symmetric formulation of the Schwarzschild initial data corresponds in the current context to the mapping
\be
\chi \rightarrow \pi -\chi.
\ee
As the sphere at $\chi=\pi/2$ is invariant under this mapping, and as the mapping is involutive, one can think of it as a reflection of the geometry in the sphere $\chi=\pi/2$. The mass point that appears to exist at $\chi=0$ is then replaced by the one at $\chi=\pi$, and {\it vice versa}. This is, of course, the same reflection symmetry in the horizon that is discussed by Misner in the asymptotically flat formulation of the same problem \cite{Misner63}. Just as in that case, the reflection symmetry shows that the initial geometry is non-singular everywhere, as indicated by the embedding diagram illustrated in Fig. \ref{zerospheres}. 

\subsection{The N-Body Problem}

Now consider adding a non-gravitating reference 2-sphere into the geometry described by Eqs. (\ref{geom}) and (\ref{nbody}). The radius of this sphere is taken to be $\chi_s=a$, where $\chi_s$ is a coordinate that is obtained by rotating ($\chi$, $\theta$, $\phi$)  into ($\chi_s$, $\theta_s$, $\phi_s$), such that the centre of the sphere is located at $\chi_s=0$. One can then define a reflection operator $J$ that operates on $\chi_s$ to give
\be
\label{jop}
\tan (\chi_s /2) \rightarrow  \tan (J \chi_s /2) = \frac{\tan^2 (a/2)} {\tan (\chi_s/2)} .
\ee
The line-element (\ref{geom}) then transforms as
\bea
\label{J1}
\hspace{-40pt} dl^2 \rightarrow& \psi^4 \left(  J \chi_s, \theta_s, \phi_s \right) \left( d (J \chi_s)^2 +\sin^2 (J \chi_s) d \theta_s^2 + \sin^2 (J \chi_s) \sin^2 \theta_s d \phi_s^2 \right)\\
&= \left[ \sqrt{\frac{\sin^2 (a/2) \cos^2 (a/2)}{{ \cos^4 (a/2) \sin^2 (\chi_s/2) + \sin^4 (a/2) \cos^2 (\chi_s/2)}}} \psi \left(  J \chi_s, \theta_s, \phi_s \right) \right]^4 d\sigma_s^2 \nonumber \;,
\eea
where $d\sigma_s^2 = d \chi_s^2 +\sin^2 \chi_s d \theta_s^2 + \sin^2 \chi_s \sin^2 \theta_s d \phi_s^2$.

We therefore define an operator $J$ that acts on functions $f=f(\chi_s, \theta_s, \phi_s)$ such that
\be
\label{jmap}
f (\chi_s, \theta_s, \phi_s) \rightarrow J [f] (\chi_s, \theta_s, \phi_s)
\ee
where
$$
J [f] (\chi_s, \theta_s, \phi_s) = \sqrt{\frac{\sin^2 (a/2) \cos^2 (a/2)}{{ \cos^4 (a/2) \sin^2 (\chi_s/2) + \sin^4 (a/2) \cos^2 (\chi_s/2)}}}\;\; f (J \chi_s, \theta_s, \phi_s).
$$
It is clear that as both a mapping (\ref{jop}) and an operator (\ref{jmap}) that $J$ satisfies the involution condition $J^2 = 1\!\!1 $, and that $J a= a$. The effect of $J$ is therefore to produce an image of the region $\chi_s >a$ in the region $\chi_s<a$, and/or and image of the region $\chi_s < a$ in the region $\chi_s >a$. These images automatically satisfy the constraint equation (\ref{constraint1b}), as $J$ has been defined such that $\nabla^2 f = \frac{3}{4} f$ implies $\nabla^2 J[f] = \frac{3}{4} J[f]$, and as Eq. (\ref{helmholtz}) is linear.

Let us now consider the effect of $J$ on the each of the terms in Eq. (\ref{nbody}). For a mass at $\chi_s=\chi_0$ it gives
\be
\label{jmass}
J \left[ \frac{\sqrt{\tilde{m}_i}}{2 \sin (\chi_i/2)} \right] = \frac{\sin a}{\sqrt{2- \sin^2 a-2 \cos a \cos \chi_0}} \; \frac{\sqrt{\tilde{m}_i}}{2 \sin (J \chi_i/2)} \; ,
\ee
where $J \chi_i$ represents the $\chi$ coordinate after a rotation of the set ($\chi$, $\theta$, $\phi$) so that $\chi =0$ at the the point $J \chi_0$ (i.e. after rotating so that the reflection of the point at $\chi_s = \chi_0$ is located at $\chi=0$). We can interpret Eq. (\ref{jmass}) as $J$ acting as the following operator on the effective mass parameter:
\be
\label{jmass2}
\tilde{m}_i \rightarrow J \tilde{m}_i = \frac{\sin^2 a}{(2- \sin^2 a-2 \cos a \cos \chi_0)} \tilde{m}_i.
\ee
The new parameter $J \tilde{m}_i$ should be considered to be the effective mass of the image point at $J \chi_0$. Any mass point originally located in the region $\chi_s >a$ can now be assigned an image mass in the region $\chi_s <a$, and any mass point originally in the region $\chi_s <a$ can be assigned an image mass in the region $\chi_s>a$. Any mass located on the sphere $\chi_s=a$ is invariant under the reflections discussed above, and so is its own image.

After taking an image of each of the masses in the space in this way one ends up with a geometry that is invariant under the reflections described above. The conformal factor can then be seen to satisfy the following equation:
\be
J[\psi]= \psi.
\ee
This means that the 2-sphere at $\chi_s=a$ is a reflection symmetric surface. As such it is totally geodesic, and an extremal surface in the initial data \cite{eisenhart}. It therefore corresponds to a MOTS, as time-reflection symmetry implies that  all extremal surfaces in the initial data are marginally outer trapped \cite{gibbons72}. Such a surface is not necessarily the MOTS of any single mass point, but can instead encompass any number of masses (each of which will have its own MOTS). In this sense it appears as an outer horizon, which are known to occur when two or more mass points cluster close enough together. Due to the reflection symmetry, however, it must simultaneously be the outer horizon of all of the masses in the region $\chi_s<a$ as well as the outer horizon of all the points in the region $\chi_s>a$. This is reminiscent of the reflection symmetric surface in the time-symmetric initial data for the Schwarzschild solution, which is simultaneously the horizon for each of the two antipodal masses.

\begin{figure}[t]
\begin{centering}
\includegraphics[width=15cm]{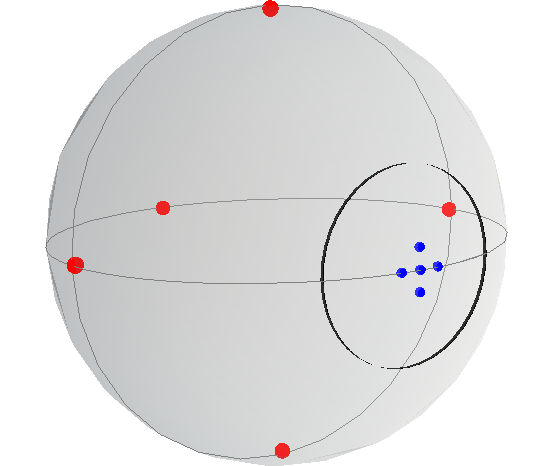}
\par\end{centering}
\caption{A dimensionally reduced illustration of the application of the method of images to mass points on a reference sphere. The red dots are unreflected mases, the black circle denotes the position of a reflective sphere, and the blue dots denote the images of the red dots. The image points are clustered around the centre of the reflecting sphere.}
\centering{}\label{sphere}
\end{figure}

\begin{figure}[t]
\begin{centering}
\includegraphics[width=9.8cm]{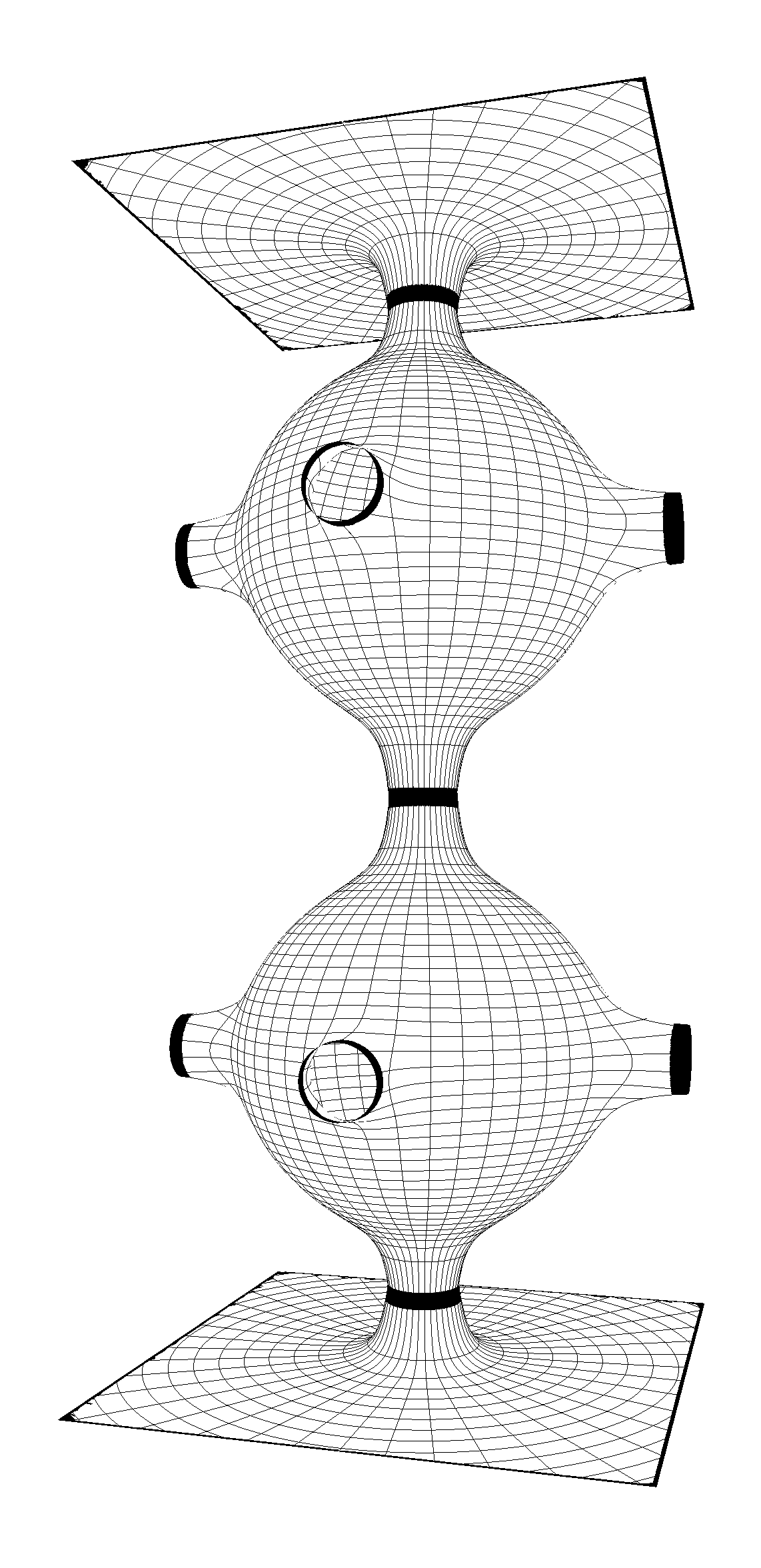}
\par\end{centering}
\caption{The dark bands in this plot represent MOTS. The mass points in the upper sphere can be considered the images of those in the lower sphere (or {\it vice versa}). In this case the reflecting sphere is at the MOTS in the central throat. The position and mass of these image points are given by Eqs. (\ref{jop}) and (\ref{jmass2}).}
\centering{}\label{twospheres}
\end{figure}

A dimensionally-reduced version of this type of reflection is shown in Fig. \ref{sphere}. Here the red dots correspond to the original mass points, which are regularly spaced on the reference sphere. A reflective sphere is added, and the positions of blue image points are calculated using Eq. (\ref{jop}). There is one image for each original mass point, and, when viewed as points on the reference sphere, the images points can be seen to be clustered around the centre of the reflective sphere. After performing a reflection of this kind there is no longer a single cosmological region and $N$ flange regions, there is instead two cosmological regions and $2 N$ flange regions (assuming there are originally no mass points inside, or on, the reflective sphere). This situation is illustrated in Fig. \ref{twospheres}. A further reflection in a different reflective sphere will then give four cosmological regions and $4 N$ flange regions (under the same assumption). This can continue indefinitely, and can potentially be used to construct wormholes by alternately reflecting in the same two spheres. Such structures will be used below to remove pairs of flange regions.

%%%%%%%%%%%%%%%%%%%%%%%%%%%%%
%%%%%%%%%%%%%%%%%%%%%%%%%%%%

%%%%%%%%%%%%%%%%%%%%%%%%%%%%%
%%%%%%%%%%%%%%%%%%%%%%%%%%%%

\section{Reflecting an 8-Black-Hole Universe}

As a concrete application of the ideas discussed in the previous section, we will now perform some reflections in the 8- black-hole universe studied in \cite{CRT12,BK, CRTG13}. This structure corresponds, in some sense, to the simplest of the six possible regular lattices that can be constructed on a 3-sphere with regular polyhedra. It can be realised by using either a lattice constructed from eight cubic lattice cells with a point-like mass positioned at the centre of each, or equivalently by using a lattice constructed from sixteen tetrahedral lattice cells and placing a point-like mass at each of the eight vertices. The coordinates of these eight points can be chosen as in Table \ref{table1}, and result in a geometry given by Eq. (\ref{nbody}) with the $f_i$ also in Table \ref{table1}. To maintain the regularity of the resultant geometry, we choose $\tilde{m}_i=\tilde{m}=$constant for each of the $8$ masses.

\begin{table}[h!]
\begin{center}
\begin{tabular}{|c|l|l|}
\hline
\bf{Point}  & \bf{($\chi$, $\theta$, $\phi$)} &  $f_i(\chi, \theta, \phi)$
\\  \hline
$(i)$ &  $\left(0, \frac{\pi}{2}, \frac{\pi}{2} \right)$  &   $f_1 = \sin \left[\frac{\chi}{2}\right] $
\\
$(ii)$ & $\left(\pi, \frac{\pi}{2}, \frac{\pi}{2} \right)$ &  $f_2 = \cos \left[\frac{\chi}{2}\right] $
\\
$(iii)$ &  $\left( \frac{\pi}{2}, 0, \frac{\pi}{2} \right)$ &   $f_3 = \sin \left[ \frac{1}{2} \cos^{-1} \left( \cos \theta
  \sin \chi \right) \right] $
\\
$(iv)$ &  $\left( \frac{\pi}{2}, \pi, \frac{\pi}{2} \right)$ &  $f_4 = \cos \left[ \frac{1}{2} \cos^{-1} \left( \cos \theta
  \sin \chi \right) \right] $
\\
$(v)$ &  $\left( \frac{\pi}{2}, \frac{\pi}{2}, 0 \right)$ &   $f_5 = \sin \left[ \frac{1}{2} \cos^{-1} \left( \cos \phi \sin \theta 
  \sin \chi \right) \right] $
\\
$(vi)$ &  $\left( \frac{\pi}{2}, \frac{\pi}{2}, \pi \right)$  &  $f_6 = \cos \left[ \frac{1}{2} \cos^{-1} \left( \cos \phi \sin \theta
  \sin \chi \right) \right] $
\\
$(vii)$ &  $\left( \frac{\pi}{2}, \frac{\pi}{2}, \frac{\pi}{2}\right)$ &  $f_7 = \sin \left[ \frac{1}{2} \cos^{-1} \left( \sin \phi \sin \theta
  \sin \chi \right) \right] $
\\
$(viii)$ &  $\left( \frac{\pi}{2}, \frac{\pi}{2}, \frac{3 \pi}{2} \right)$ & $f_8 = \cos \left[ \frac{1}{2} \cos^{-1} \left( \sin \phi \sin \theta 
  \sin \chi \right) \right]$
\\ \hline
\end{tabular}
\end{center}
\caption{{\protect{\textit{Coordinates of the eight masses in the 8-black hole universe, written in hyperspherical polar coordinates. These coordinates are chosen such that point $(i)$ appears at $\chi=0$, point $(iii)$ appears at $\theta=0$, and point $(v)$ appears at $\phi=0$. Also displayed are the functions from Eq. (\ref{nbody}) for these eight mass points, in the same coordinate system. Note that $cos^{-1}$ denotes the inverse of the cosine, and not its reciprocal.}}}}
\label{table1}
\end{table}

\subsection{One Reflection}

The first application of the method of images we wish to consider is a single set of reflections in a sphere that is centred around the point $\chi=0$, and that has its surface at $\chi=a$. This central point is coincident with the position of mass $(i)$, as described in Table \ref{table1}. Our method is then given by the following: Firstly, we discard the point $(i)$, so that there are initially no masses inside the sphere. We then calculate the locations and masses of the images of the $7$ masses that remain in the space. This is done using Eqs. (\ref{jop}) and (\ref{jmass2}). We then determine the geometry of this new space using Eq. (\ref{nbody}), and hence calculate the proper masses and horizon areas of the black holes after the reflection (recall that these will be different to those before the reflection, as they are functions of all of the masses in the universe). The embedding diagram for a $2$-dimensional slice through this geometry will look like that displayed in Fig. \ref{twospheres}.

\begin{figure}[t]
\begin{centering}
\includegraphics[width=15cm]{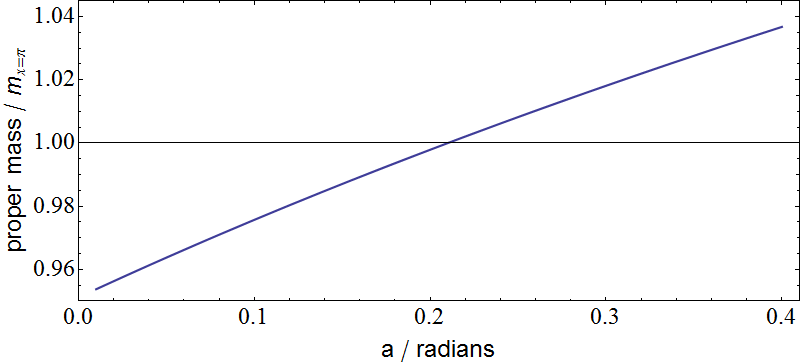}
\par\end{centering}
\caption{The proper mass of the black hole located at position $(ii)$ in Table \ref{table1}, as a proportion of the proper mass of any of the $6$ black holes at positions $(iii)$-$(viii)$. The critical value of $a$ is given when this ratio is equal to $1$.}
\centering{}\label{massplot}
\end{figure}

The proper mass of each of the new image black holes is guaranteed to be the same as that of the black hole from which the image was taken. This follows immediately from Eqs. (\ref{J1}) and (\ref{jmap}), and has been verified explicitly for the example in hand. This is not sufficient, however, to say that all black holes have the same proper mass after a reflection has taken place, even if they had the same proper mass in the original 8-black hole universe. In general, after a reflection, the black hole at point $(ii)$ will not have the same proper mass as any of the masses at points $(iii)$-$(viii)$ (although all of the black holes at points $(iii)$-$(viii)$ will be the same, by symmetry). The difference in proper mass between these objects is shown graphically in Fig. \ref{massplot}. It can be seen that if the reflective sphere is smaller than a critical value of $a$ then the proper mass at point $(ii)$ will be smaller than the proper mass of any of the $6$ objects at points $(iii)$-$(viii)$. This critical value is given by
\be
\label{acrit}
a_{\rm crit} \simeq 0.2100769668987814 \; .
\ee
For $a > a_{\rm crit}$ the opposite is true, and the proper mass at point $(ii)$ is larger. If $a=a_{\rm crit}$ then all $14$ masses that exist after the reflection have identical proper masses.

%%%%%%%%%%%%%%%%%%%%%%%%%%%%%
%%%%%%%%%%%%%%%%%%%%%%%%%%%%

\begin{figure}[t]
\begin{centering}
\includegraphics[width=15cm]{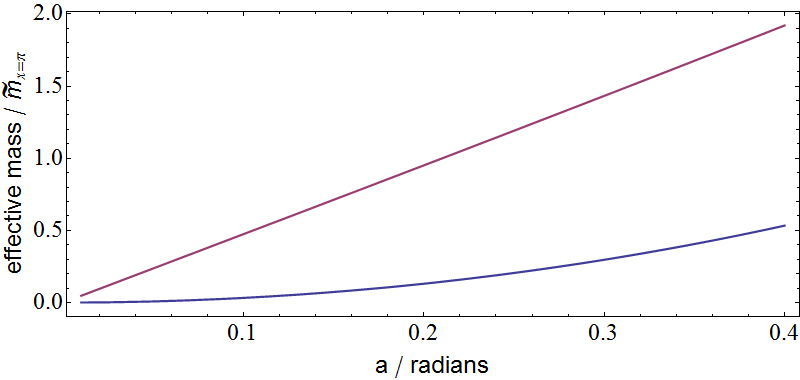}
\par\end{centering}
\caption{The sum of effective masses in the region $\chi < a$, after a reflection has taken place (lower, blue line). Also displayed is the sum of the charges in the same region (upper, red line).}
\centering{}\label{massplot2}
\end{figure}

As well as the proper mass of each of the black holes, it is also of interest to calculate the effective mass that results from the reflection described above. This information is displayed in Fig. \ref{massplot2}, below. This plot illustrates the total effective mass of all of the new image black holes, as a function of the radius of the reflective sphere, $a$. One may naively have expected this quantity to be close to $1$, but it can be seen that total effective mass inside the reflective sphere is in fact $<1/2$ of the mass they replaced, for all $a \lesssim 0.4$. Also displayed in this plot is the total of the charges of all of the images. This quantity can be seen to change rapidly, but at $a=a_{\rm crit}$ it takes the value 
\be
\label{1charge}
\sum_i \sqrt{\tilde{m}_i} \vert_{a=a_{\rm crit}} \simeq 0.99994472 \, \sqrt{{\tilde m}_{\chi=\pi}} \;.
\ee
This is remarkably close to $1$, and can be compared to the value of the total effective mass of these points at the same point,
$
\sum_i \tilde{m}_i \vert_{a=a_{\rm crit}} \simeq 0.14447522 \, {\tilde m}_{\chi=\pi} \;.
$

%%%%%%%%%%%%%%%%%%%%%%%%%%%%%
%%%%%%%%%%%%%%%%%%%%%%%%%%%%

We can now calculate the proper area of our reflective sphere (which is a MOTS, as described above). If we do this for a sphere of radius $a=a_{\rm crit}$, and compare it to the area of a Schwarzschild black hole with the same proper mass of any of the $14$ black holes in the space, then we find that the fractional difference is
$$
\frac{ \delta {\rm area}}{{\rm area}} \simeq 5.4738620 \times 10^{-8} \; .
$$ 
We can similarly investigate the length of one of the edges of our cubic lattice cells. For this we choose an edge that has constant $\theta$ and $\phi$ at every point along it. The fractional difference in this quantity before and after the reflection, again for a reflective sphere of radius $a=a_{\rm crit}$, is given by
$$
\frac{\delta {\rm edge}}{{\rm edge}} \simeq -6.2648829 \times 10^{-9} \; .
$$
We therefore find that, for a reflective sphere of critical radius, both the horizon size and the scale of our cosmological region are very similar before and after the reflection has taken place, with changes of only around $1$ part in $10^8$.

This result is interesting, as the positions of the $7$ image masses on the reference hypersphere all appear to be clustered into a relatively small space (i.e. are all separated by $\chi \ll 1$). They could therefore be considered a cluster of masses, as discussed in Section \ref{sec:energy}. The results we have found then show that the gravitational field of this cluster is similar to the single mass they replace, if $a \simeq a_{\rm crit}$. They also show that the total charge of this cluster of masses is very similar to the charge of the single mass they have replaced (as shown in Eq. (\ref{1charge})). This suggests that the quantity $M$ that we introduced in Eq. (\ref{M}) is indeed a good approximation to the gravitational mass of a cluster, and that the interaction energies between the masses in this cluster {\it do} appear to contribute to the scale of the cosmological region, at least in the present case. This is true because $B$ in Eq. (\ref{M2}) is not affected by the reflection.

\subsection{n-Reflections of a Set of Identical Black Holes}

In the previous section we discussed a single reflection in a sphere centred around point $(i)$ of Table \ref{table1}. It was found that the proper mass of the resultant $14$ black holes were equal to each other only if the sphere was chosen to have a radius of $a=a_{\rm crit}$ (given numerically in Eq. (\ref{acrit})). In this section we will consider a series of reflections in two spheres. The first of these spheres will be centred around point $(i)$, as before. The second sphere will be centred around point $(ii)$, which is located at $\chi=\pi$. These two points are antipodal from each other, and we will now consider a series of reflection operations that reflect in these two spheres alternately ({\it i.e.} first in a sphere centred on $\chi=0$, then in a sphere centred on $\chi=\pi$, then in a sphere centred at $\chi=0$, {\it etc.}).

%%%%%%%%%%%%%%%%%%%%%%%%%%%%%
\begin{figure}[t]
\begin{centering}
\includegraphics[width=15cm]{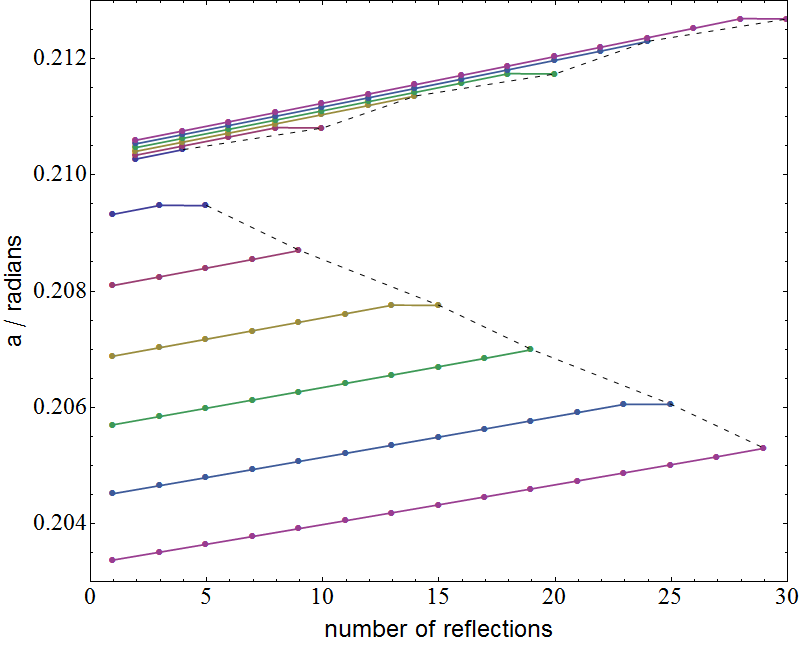}
\par\end{centering}
\caption{The radius of the reflective spheres required to keep the proper mass of every black hole the same after a series of $5$ (dark blue), $10$ (red), $15$ (yellow), $20$ (green), $25$ (light blue), and $30$ (pink) reflections. In each series the spheres centred at $\chi=0$ are connected by solid lines, as are spheres centred at $\chi=\pi$. Dashed lines show the trend for the radius of the final sphere at each antipodal point, as the number of reflections in the series is increased.}
\centering{}\label{massplot3}
\end{figure}
%%%%%%%%%%%%%%%%%%%%%%%%%%%%

Just as in the case of a single reflection, the creation of image points will in general change the proper mass of all other black holes that exist in the space. Just as before, however, there is a way that we can choose the size of each of the reflecting spheres such that the proper mass of all black holes remain equal. That this is possible can be seen by considering that after $n$ reflections we will have a total of $12 {n} + 2$ black holes. That is there will be $2 n$ sets of $6$ black holes, plus $1$ black hole at $\chi=0$, plus $1$ more at $\chi=\pi$. Each black hole in each set of $6$ must have the same proper mass by symmetry, as must the two black holes at $\chi=0$ and $\chi=\pi$ (as they are always images of each other). This leaves $2 n+1$ sets of black holes that can have different masses from each other. However, after each reflection, half of the $2 n$ sets of $6$ black holes must contain black holes that have the same proper mass as those in one of the sets in the other half, as they are images. After $n$ reflections we therefore have $n+1$ sets of black holes, each of which can in principal contain black holes that have different masses from those in each of the other sets. We therefore have enough freedom in the radii of our $n$ reflective spheres to set the proper masses of all of our black holes to be equal, while retaining the freedom to set this single proper mass to any value we choose. 

%%%%%%%%%%%%%%%%%%%%%%%%%%%%%
\begin{figure}[t]
\begin{centering}
\includegraphics[width=15cm]{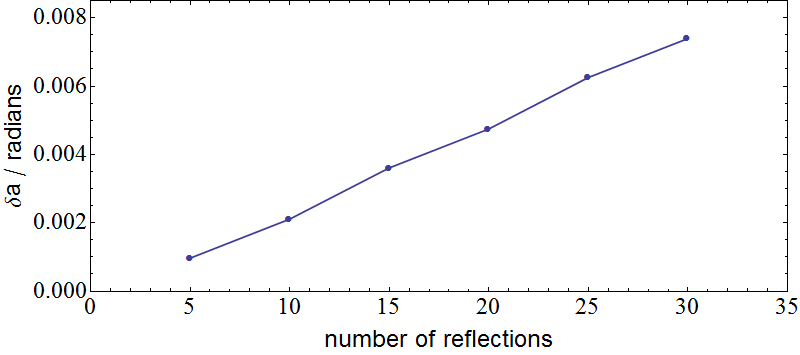}
\par\end{centering}
\caption{The difference in radii of the final two reflecting spheres, as a function of the number of reflections in the series.}
\centering{}\label{massplot4}
\end{figure}
%%%%%%%%%%%%%%%%%%%%%%%%%%%%

It can be noted that the method just described will require us to change the radius of the each of the reflective spheres after every reflection in the series. That is, if we want all of our black holes to have identical proper mass, we cannot simply place a single sphere with radius $\chi=a_1$ at point $(i)$, and a single sphere of radius $\chi=a_2$ at point $(ii)$, and then reflect alternately in these two spheres. We must instead allow ourselves the freedom to choose the radius of each of the reflecting spheres of each reflection event individually. The results of doing exactly this are displayed in Fig. \ref{massplot3}. In this figure we have plotted the results of considering six different series, that have $5$, $10$, $15$, $20$, $25$ and $30$ reflections in each of them, respectively. The $y$-axis shows the radius that the reflecting sphere of each reflection event must have in order for every black hole in the space to have the same proper mass after the entire series of reflections is complete. In every case the first reflection in taken in a sphere centred at $\chi=0$.

%%%%%%%%%%%%%%%%%%%%%%%%%%%%%
\begin{figure}[t]
\begin{centering}
%\vspace{1.5cm}
\includegraphics[width=15cm]{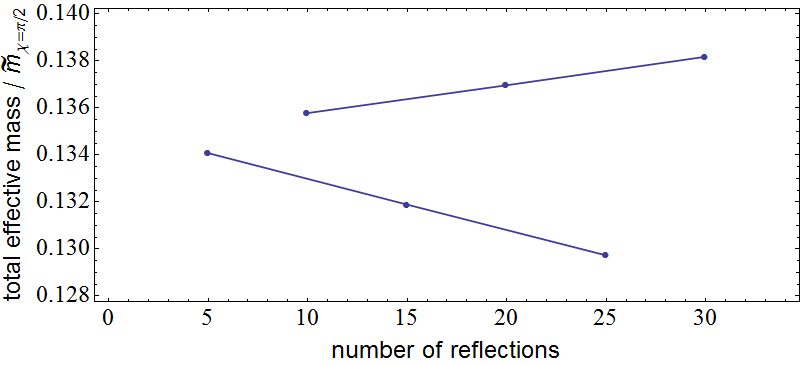}
\par\end{centering}
\caption{The total effective mass in the final reflecting sphere, as a function of the number of reflections in the series, and as a fraction of the effective mass of one of the black holes at $\chi=\pi/2$.}
%\vspace{2cm}
\centering{}\label{massplot5}
\end{figure}
%%%%%%%%%%%%%%%%%%%%%%%%%%%%

It can be seen from Fig. \ref{massplot3} that, within each series, the trend for spheres centred around $\chi=0$ is to increase as the reflections are performed. That is, to have identical black holes after a series is complete, we need the reflecting spheres around $\chi=0$ to be smaller for the first reflection than they are for every subsequent (even numbered) reflection. A similar trend can be seen for the spheres centred around $\chi=\pi$. It can also be seen, however, that {\it all} of the reflecting spheres centred around $\chi=\pi$ must be smaller than {\it all} of the spheres centred around $\chi=0$, in every series considered. This broken symmetry is allowed, as the choice to start each series with a reflection in the sphere centred on $\chi=0$ itself breaks symmetry. A further pattern can be seen as the length of the series is increased. In this case, the radius of the spheres centred around $\chi=0$ increase, while those centred around $\chi=\pi$ decrease. This phenomenon is further illustrated in Fig. \ref{massplot4}, where the difference in the radii of the final two reflective spheres is plotted as a function of the number of reflections in the series. The difference increases approximately linearly with the number of reflections.

\begin{figure}[t]
\begin{centering}
\includegraphics[width=15cm]{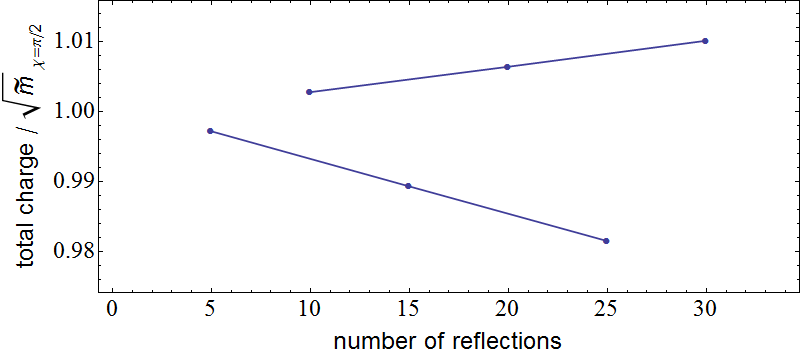}
\par\end{centering}
\caption{The total charge in the final reflecting sphere, as a function of the number of reflections in the series, and as a fraction of the charge of one of the black holes at $\chi=\pi/2$.}
%\vspace{1cm}
\centering{}\label{massplot6}
\end{figure}

Finally, we can calculate the total mass inside each of the spheres, for each of the series we have so far considered. The total effective mass in the last sphere in each series is displayed in Fig. \ref{massplot5}. It can be seen from this figure that the total effective mass increases if the total number of reflections is even, and decreases if the total number of reflections is odd. This dependence on whether the number of reflections is even or odd can be understood from the different behaviour of the spheres centred around $\chi=0$ and $\chi=\pi$ in Fig. \ref{massplot3}. Similarly, in Fig. \ref{massplot6} we show the total charge in the final sphere of each of our series of reflections. The total charge can also be seen to increase or decrease, depending on whether the number of reflections in the series is even or odd. It can also be seen to diverge away from unity. The result from Eq. (\ref{1charge}), that the numerical value of the charge after one reflection is very close to $1$, does not therefore appear to hold if a large number of reflections are performed. Instead it can be seen that after $\sim 30$ reflections the difference in charge is at the level of $\sim 1$ or $2\%$. We will comment on the significance of this, with regard to the cosmological consequences of interaction energies, in Section \ref{discussion}.

%%%%%%%%%%%%%%%%%%%%%%%%%%%%%
%%%%%%%%%%%%%%%%%%%%%%%%%%%%

It is clear from the above that if we insist that the proper mass of every black hole should be identical to every other, then there is no sign of convergence of either the effective mass, the charge, or the radii of the reflecting spheres (for series of up to $30$ reflections, at least). This means that the space itself is also not converging on any particular geometry. In the next section we will discard the idea that all black holes should have the same proper mass, and will find that it is then possible to find convergence as the number of reflections in a series is increased. The existence of this convergence will allow us to construct wormholes, and hence remove some of the flange regions that exist in the original 8-black hole universe.

\subsection{A Convergent Series of Reflections}

To find a series of reflections that converging, let us consider a sum of poles of the form
\be
\label{psisum}
\psi =  \sum_i \frac{q_i}{2 \sin (\chi_i/2)},
\ee
where $q_i$ are a set of constant charges. Now consider a region $R$ that is bounded away from the poles, so that we have $\sin (\chi_i/2) \geq \sin (\rho /2) >0$ for all $i$. This gives
\be
\label{psiseries}
\sum_i \frac{q_i}{2 \sin (\chi_i/2)} \leq \sum_i \frac{\vert q_i \vert}{2 \sin (\chi_i/2)} \leq \frac{1}{2 \sin(\rho/2)} \sum_i \vert q_i \vert \; ,
\ee
which shows that $\psi$ is finite in $R$ if $\sum_i \vert q_i \vert$ is finite.

As before, let us consider putting a reflective sphere around $\chi=0$ and a second sphere around $\chi = \pi$. This time let us consider both of these spheres to have a constant and fixed radius of $\chi = a$. Once again, we will begin by reflecting in the sphere centred on $\chi=0$, then reflecting in the sphere centred on $\chi=\pi$. We will then reflect in the sphere centred on $\chi=0$ again, and the sphere centred on $\chi=\pi$ again. This will be continued {\it ad infinitum}.

Let us consider the series of images that are collected in the sphere centred on $\chi =0$. The first images are all at $\chi = 2 \tan^{-1} \left[ \tan^2 (a/2) \right]$, and there are six of them (one for each of the mass points that is not enclosed by a sphere). The next set of images in this sphere are all at $\chi = 2 \tan^{-1} \left[ \tan^4 (a/2) \right]$, and are the images of the first set of images collected in the sphere centred around $\chi=\pi$. The third set of images are then the images of the images of images. This series continues forever with the $i$th set of images being located at
\be
\chi_i = 2 \tan^{-1} \left[ \tan^{2 i} (a/2) \right].
\ee
Likewise, the charge of each of these sets of images can be readily calculated, and is given by
\be
q_i = \sqrt{\tilde{m}} \; \prod_{j=1}^i \frac{\sin a}{\sqrt{2-\sin^2 a + 2 \cos a \cos \left( 2 \tan^{-1} \left( \tan^{2 (j-1)} (a/2) \right)  \right)}} \; .
\ee
The total charge in this sphere is therefore given by
\be
\label{qseries}
6 \sum_{i=1}^{\infty} q_i.
\ee
To see if this series is convergent we can calculate the ratio of two successive terms to find
\be
\frac{q_{i+1}}{q_i} = \frac{\sin a}{\sqrt{2-\sin^2 a + 2 \cos a \cos \left( 2 \tan^{-1} \left( \tan^{2 i} (a/2) \right)  \right)}}.
\ee
In the limit $i \rightarrow \infty$ we then have
\be
\lim_{i \to \infty} \left\vert \frac{q_{i+1}}{q_i} \right\vert = \frac{\sin a}{\sqrt{2- \sin^2 a + 2 \cos a}}.
\ee
This quantity is less than one if $0 < a < \pi/2$, and so the series in Eq. (\ref{qseries}) converges by D'Alembert's criterion. From Eq. (\ref{psiseries}) we then have that the contribution to $\psi$ from the images in the sphere centred on $\chi=0$ is convergent if $0 < a < \pi/2$. The same is true of the sphere centred on $\chi=\pi$ by symmetry, and so the series defined in Eq. (\ref{psisum}) is convergent if $0 < a < \pi/2$.

%%%%%%%%%%%%%%%%%%%%%%%%%%%%%
\begin{figure}[t]
\begin{centering}
\includegraphics[width=15cm]{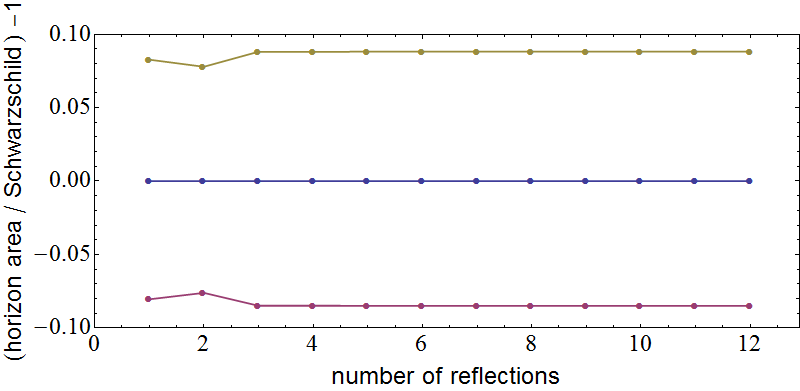}
\par\end{centering}
\caption{The proper area of the last reflecting sphere as a fraction of the horizon area of a Schwarzschild black hole with the same mass as one of the unreflected black holes at $\chi=\pi/2$, after a given number of reflections in alternating antipodal points. The Blue line is with $a=a_{\rm crit}$, the red and yellow lines are with $a=a_{\rm crit}-0.01$ and $a=a_{\rm crit}+0.01$, respectively.}
\centering{}\label{Nref1}
\end{figure}
%%%%%%%%%%%%%%%%%%%%%%%%%%%%

First of all, let us consider the proper area of the final reflecting sphere, after $n$ reflections. This is shown graphically in Fig. \ref{Nref1} for spheres of radius $\chi=a_{\rm crit}$, and $\chi = a_{\rm crit} \pm 0.01$ (the numerical value of $a_{\rm crit}$ is given in Eq. (\ref{acrit}). It can be seen that in all three of these cases the area of the horizon converges to a constant value after only a small number of reflections. Such a convergence should be expected from the discussion above, as it is a direct function of $\psi$, which has proven to be convergent. It can be seen that spheres of critical radius, which are already known to correspond to MOTSs, have an area that is very close to the area of a Schwarzschild black hole with a mass equal to the effective mass of one of the unreflected black holes at $\chi=\pi/2$. Increasing the size of the reflecting sphere increases the area of the MOTS, and decreasing it does the opposite.

\begin{figure}[t]
\begin{centering}
\includegraphics[width=15cm]{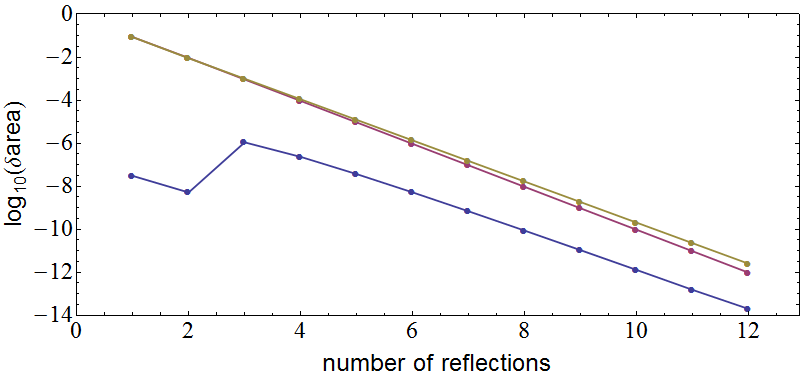}
\par\end{centering}
\caption{The logarithm of the fractional difference in area of the antipodal reflecting spheres after a given number of reflections. Lines as in Fig. \ref{Nref1}.}
\centering{}\label{Nref2}
\end{figure}

We can also calculate the fractional difference in area between the last two reflecting spheres, for a given series of reflections. In the previous section we showed that this diverges if the proper mass of all black holes are forced to be equal (see Fig. \ref{massplot4}). Here this is not the case, as can be seen from Fig. \ref{Nref2}. After a small amount of irregularity for low numbers of reflections, it can be seen that the area of the sphere around $\chi=\pi$ rapidly approaches that of the sphere around $\chi=0$. In fact, the fractional difference in area between these two spheres decreases exponentially as the number of reflections increases. Again, this is not surprising, and follows from the convergence of $\psi$ as the number of reflections diverges.

%%%%%%%%%%%%%%%%%%%%%%%%%%%%%
%%%%%%%%%%%%%%%%%%%%%%%%%%%%

\begin{figure}[t]
\begin{centering}
\includegraphics[width=15cm]{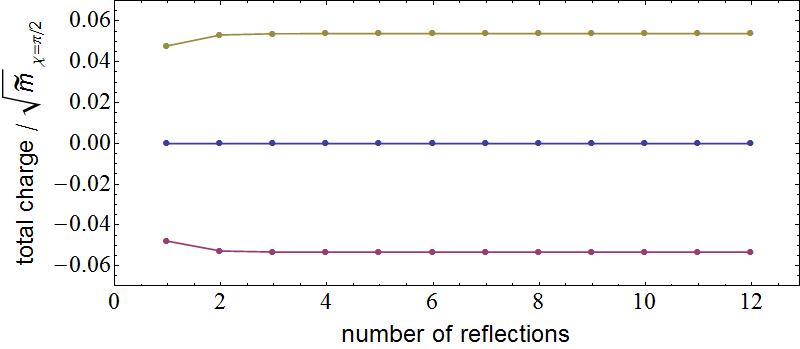}
\par\end{centering}
\caption{The total charge in the last reflecting sphere after a given number of reflections.  Lines as in Fig. \ref{Nref1}.}
\centering{}\label{Nref3}
\end{figure}

Finally, let us consider the charge and proper mass of the image points in the final reflecting sphere. The total charge is plotted in Fig. \ref{Nref3}, for the same three values of $a$ considered in Figs. \ref{Nref1} and \ref{Nref2}. As with the proper area of the spheres, it can be seen that the total charge within the final reflective sphere rapidly converges as the number of reflections is increased (this can be compared with the contrasting behaviour displayed in Fig. \ref{massplot6}). Increasing the radii of these spheres can be seen to correspondingly increase the total charge of the image points within them. Decreasing the area does the opposite. This explicitly demonstrates the convergence required in Eq. (\ref{psiseries}).

%%%%%%%%%%%%%%%%%%%%%%%%%%%%%
%%%%%%%%%%%%%%%%%%%%%%%%%%%%

\begin{figure}[t]
\begin{centering}
\includegraphics[width=15cm]{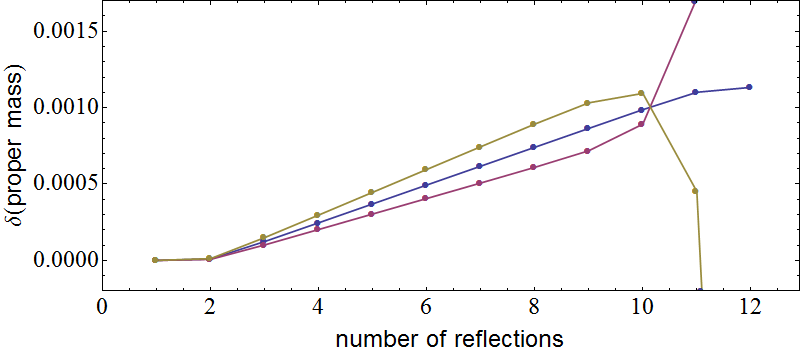}
\par\end{centering}
\caption{The fractional difference between the proper masses of points in the last reflecting sphere, and the points that have never been reflected, after $12$ reflections.  Lines as in Fig. \ref{Nref1}.}
\centering{}\label{Nref4}
\end{figure}

\begin{figure}[t]
\begin{centering}
\includegraphics[width=15cm]{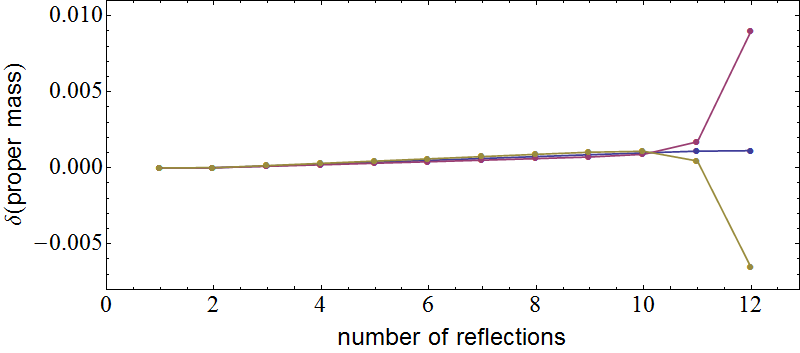}
\par\end{centering}
\caption{The same as in Fig. \ref{Nref4}, but zoomed out so that all points are displayed.  Lines as in Fig. \ref{Nref1}.}
\centering{}\label{Nref5}
\end{figure}

The proper mass of each of the image points is shown graphically in Figs. \ref{Nref4} and \ref{Nref5}, where we plot the fractional difference in proper mass of each of the image masses when compared to the original masses. The $x$-axis in these plots corresponds to the number of times a particular image point has been reflected. Fig. \ref{Nref4} shows that images that have been reflected only once or twice have proper masses that are very similar to the unreflected masses, while images that have been reflected many times can have quite different values. This difference in proper mass increases with the number of reflections, and is a function of the size of the reflecting sphere. For a sphere of critical radius, the fractional difference in proper mass of images that have been reflected either $10$ or $12$ times is at the level of about $1\%$. For spheres of radius $a=a_{\rm crit} \pm 0.01$ this difference is about $1\%$ after 10 reflections, and increases to more than $5\%$ after 12 reflections (see Fig. \ref{Nref5}). This shows that while the charge in the reflective sphere converges, and hence so does the value of $\psi$ in the region bounded away from the singularities, the proper mass of the black holes in the most distant cosmological regions becomes more and more different from the proper mass of the black holes in the original cosmological region.

%%%%%%%%%%%%%%%%%%%%%%%%%%%%%
%%%%%%%%%%%%%%%%%%%%%%%%%%%%

%%%%%%%%%%%%%%%%%%%%%%%%%%%%%
%%%%%%%%%%%%%%%%%%%%%%%%%%%%

\subsection{Six Flanges, and One Wormhole}
\label{sec:1wh}

Let us now use this convergent series of reflections to construct a wormhole. To do this we can write down the form of $\psi$ as the number of reflections diverges:
\be
\label{psiinfty}
\psi_{\infty} \equiv \sum_{j=1}^6 \frac{\sqrt{\tilde{m}_j}}{2 \sin (\chi_j/2)}+ \lim_{N \rightarrow \infty} \sum_{n=1}^N  \sum_{j=1}^6   J_{i_1} J_{i_2} ... J_{i_n}  \left[   \frac{\sqrt{\tilde{m}_j}}{2 \sin (\chi_j/2)}  \right] \, ,
\ee
where $j$ runs from $1$ to $6$, and labels the $6$ original black holes that lie outside of both reflective spheres, and where the indices $i_k$ all run from $1$ to $2$, and label the two reflecting spheres. The $i_k$ indices are used here in such a way that $i_{k+1} \neq i_k$. It can be shown that this expression satisfies
\be
\label{JiS}
J_k \psi_{\infty} = \psi_{\infty} \, ,
\ee
where $k$ equals $1$ or $2$, and again labels the reflecting spheres. This equation states that the contribution to $\psi$ is unchanged by any further reflections of $\psi_{\infty}$. The proof of Eq. (\ref{JiS}) follows from considering the fact that reflecting any particular term in $\psi_{\infty}$ results in a term that is already included in $\psi_{\infty}$, as $\psi_{\infty}$ is the sum of all possible reflections.

\begin{figure}[t]
\begin{centering}
\includegraphics[width=10cm]{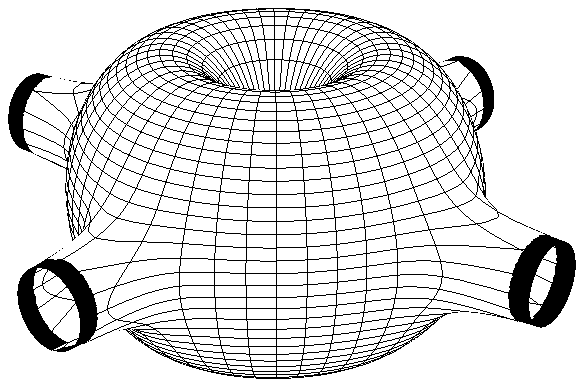}
\par\end{centering}
\caption{An illustration of the embedding diagram of a 2-dimensional space containing four MOTS, and a wormhole that connects the North and South poles. Each of the MOTS, which are denoted by thick bands, is connected to an asymptotically flat region (not included in the illustration).}
\centering{}\label{onewormhole}
\end{figure}

We now wish to show that the geometry given by Eqs. (\ref{geom}) and (\ref{psiinfty}) can be used to describe a manifold, $M$, of the type shown in Fig. \ref{onewormhole}. The discussion of how to do this will follow that devised by Misner for the asymptotically flat case \cite{Misner63}. Firstly, we need to show that each point in $M$ exists in the interior of at least one coordinate patch, in an set of overlapping coordinate systems that cover the entire space. Secondly, we need to show that the convergent geometry discussed above can be used to describe this space. This will be shown to be true if the two reflecting spheres about $\chi=0$ and $\chi=\pi$ are identified.

Begin by considering the set  of points $U$ in a region of space outside of the sphere centred at $\chi=0$, that contains the points on the sphere itself, but that does not extend far enough to include any other MOTS. Let us label these points $x$, and define a coordinate system $\chi(x)$, $\theta(x)$ and $\phi(x)$. These can be the coordinates used in Eq. (\ref{geom}). Now consider a second set of points $V$ in a region of space outside of the sphere centred at $\chi=\pi$, that does not extend far enough to reach any other MOTS, and that this time does {\it not} include the points on the sphere itself. Let us label these points $y$, and define a coordinate system $\hat{\chi}(y)$, $\hat{\theta}(y)$ and $\hat{\phi}(y)$. These coordinates could be taken to be similar to those given in Eq. (\ref{geom}), but rotated so that the centre of this second sphere appears at $\hat{\chi}=0$, and parity transformed in $\hat{\theta}$ and $\hat{\phi}$ for later convenience. A set of points $W$ that correspond to the wormhole in Fig. \ref{onewormhole} can then be defined by
\be
W = \{ z \; \vert \; {\rm either} \; z = x \in U \quad {\rm or} \quad z=y \in V \} \,.
\ee
Each point $z$ can already be seen to lie inside at least one coordinate patch, in a system of overlapping coordinates, at all points except those that exist on the sphere in $U$, which we will label $\partial U$. 

To extend the coordinate systems defined on $U$ and $V$, so that they overlap on $\partial U$, let us define a new coordinate system by
\be
\hat{\chi} (x) \equiv J \chi (x) \, .
\ee
These coordinates are regular, as the mapping defined by $J$ is regular. On $\partial U$ they satisfy $\hat{\chi}(x) = \chi (x)$, and they can be used to extend $\chi (x)$ into a region interior to the sphere centred at $\chi=0$. One can then use the expression $\hat{\chi} (x) =\hat{\chi}(y)$ to identify points $x \in U$ with new points $y$ that exist in or on the sphere centred at $\chi=\pi$. This provides us with overlapping coordinate systems on $W$ that can be regularly related to each other.

In the region covered by the $\chi$, $\theta$ and $\phi$ coordinates we have
\be
dl^2 = \psi^4 (\chi,\theta,\phi) \left(  d \chi^2 +\sin^2 \chi d \theta^2 + \sin^2 \chi \sin^2 \theta d \phi^2 \right) \, ,
\ee
and in the region covered by the $\hat{\chi}$, $\hat{\theta}$ and $\hat{\phi}$ coordinates we have
\be
dl^2 = \psi^4 (\hat{ \chi},\hat{\theta},\hat{\phi}) \left(  d \hat{\chi}^2 +\sin^2 \hat{\chi} d \hat{\theta}^2 + \sin^2 \hat{\chi} \sin^2 \hat{\theta} d \hat{\phi}^2 \right) \, .
\ee
In the regions where these coordinate systems overlap we need these two line-elements to be consistent. The coordinate transformations that relate these two sets of coordinates are given by Eq. (\ref{jop}) and $\hat{\theta}=\theta$ and $\hat{\phi}=\phi$, which on substitution gives Eq. (\ref{JiS}) as the consistency condition. The convergent geometry we discovered in the previous section can therefore be used to describe the geometry of a wormhole of the type depicted in Fig. \ref{onewormhole}.

%%%%%%%%%%%%%%%%%%%%%%%%%%%%%
%%%%%%%%%%%%%%%%%%%%%%%%%%%%

\section{A Compact Vacuum Universe}

Having created a single wormhole between points $(i)$ and $(ii)$ (as listed in Table \ref{table1}), we can now consider making a space that contains four wormholes, each of which connects a mass point with its antipodal counterpart. Such a space would have no flange regions, and would consist solely of a single compact cosmological region. We will construct just such a space in what follows.

\subsection{Convergence of Reflecting Around Eight Masses Simultaneously}

Consider the case of a reflective sphere being placed around each of the masses in the 8-black hole Universe. Each of these spheres is centred on each of the masses, and each of them is of radius $\chi_i=a$. These spheres can be used to determine an image of all the masses exterior to each of them, and to then replace the mass points interior to each of them with this image. If this is done simultaneously then one ends up with eight new cosmological regions attached to the original cosmological region by eight separate bridges. Each of these new regions contains the end to one of these bridges, and seven bridges to seven new flanges. Repeating this process $n$ times results in $1+8 \sum_{i=0}^n 7^i$ cosmological regions, and $8 \times 7^n$  flanges. Our first concern is then whether or not this is a convergent process in the limit $n \rightarrow \infty$.

Each reflection in each sphere contains an image of the collection of mass points that are contained within the antipodal sphere, and also contains six images of the mass points contained within the six spheres that are centred around points that are $\chi_i=\pi/2$ away. Let us start by considering the mass points contained within the antipodal sphere. These points are all more than $\pi-a$ radians away from the centre of the sphere in which they are to be reflected. Of the other six spheres, at least half of the mass points must be more than $\pi/2$ radians or more away, and the rest of them must be more than $\pi/2-a$ away. After the first reflection we therefore have that the total charge within each sphere must be given by the following inequality:
$$
q_1 \leq \frac{\sqrt{\tilde{m}} \sin a}{\sqrt{2-\sin^2 a +2 \cos^2 a}} +3 \frac{\sqrt{\tilde{m}} \sin a}{\sqrt{2-\sin^2 a}} + 3 \frac{\sqrt{\tilde{m}} \sin a}{\sqrt{2-\sin^2 a -2 \cos a \sin a}}.
$$
As each of the spheres must be identical after each of the reflections is performed, we have the following recursion relation, which can be used to determine an upper bound on every $q_i$:
$$
q_{i+1} \leq \frac{q_i \sin a}{\sqrt{2-\sin^2 a +2 \cos^2 a}} +3 \frac{q_i \sin a}{\sqrt{2-\sin^2 a}} + 3 \frac{q_i \sin a}{\sqrt{2-\sin^2 a -2 \cos a \sin a}}.
$$
This expression can then be used to determine the following upper bound on the difference in the charge between two successive reflections:
$$
q_{i+1} - q_i \leq 3 \frac{q_i \sin a}{\sqrt{2-\sin^2 a}} + 3 \frac{q_i \sin a}{\sqrt{2-\sin^2 a -2 \cos a \sin a}}.
$$
The total charge in each sphere after $i+1$ reflections is therefore bounded by the following expression:
$$
q_{i+1} \leq \sqrt{\tilde{m}} \left[ 3 \frac{\sin a}{\sqrt{2-\sin^2 a}} + 3 \frac{\sin a}{\sqrt{2-\sin^2 a -2 \cos a \sin a}} \right]^{i}.
$$
An infinite series of reflections therefore converges if
\be
\lim_{i \to \infty} \left\vert \frac{q_{i+1}}{q_i} \right\vert = 3 \frac{\sin a}{\sqrt{2-\sin^2 a}} + 3 \frac{\sin a}{\sqrt{2-\sin^2 a -2 \cos a \sin a}} < 1 \; .
\ee
This final inequality is satisfied if $a < 0.220316$, to six significant figures. As was the case for the limit of an infinite number of reflections in two spheres, the convergence of the total amount of charge in each sphere is enough to guarantee the convergence of $\psi$ in the same limit. This can be seen from Eq. (\ref{psiseries}). The process of reflecting all points in eight separate spheres of radius $\chi_i=a$ therefore gives a convergent sum in Eq. (\ref{psisum}) if $a<0.220316$.

\begin{figure}[t]
\begin{centering}
\includegraphics[width=15cm]{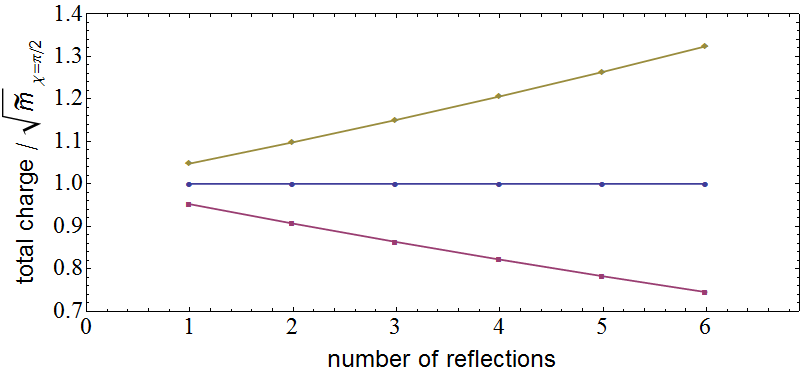}
\par\end{centering}
\caption{The total charge in one reflecting sphere after a given number of reflections of all other points. The blue line is for $a=a_{\rm flat}$, the red and yellow lines are for $a=a_{\rm flat}-0.01$ and $a=a_{\rm flat}+0.01$, respectively.}
\centering{}\label{4wormsfig1}
\end{figure}

To demonstrate this convergence explicitly we perform the reflection operation described above. The result of this up to $6$ reflection operations is displayed in Fig. \ref{4wormsfig1}. Here we have considered three different values for the radii of our reflecting spheres. The first is a choice that has been made to ensure convergence happens as rapidly as possible. This is given by:
\be
a_{\rm flat} \simeq 0.2100769710834392 \; ,
\ee
which is about $2 \times 10^{-8}$ times larger than $a_{\rm crit}$. The other two values are given by $a=a_{\rm flat} \pm 0.01$. All of these values are within the bound $a<0.220316$, which guarantees convergence. It can be seen from the figure, however, that the convergence must occur much more slowly than was the case for alternate reflections in only two antipodal spheres. Up to the six reflections displayed, the two curves with $a=a_{\rm flat} \pm 0.01$ show no signs of convergence, even though we have already shown that convergence must occur in both of these cases in the limit $n \rightarrow \infty$.

%%%%%%%%%%%%%%%%%%%%%%%%%%%%%
%%%%%%%%%%%%%%%%%%%%%%%%%%%%

\begin{figure}[t]
\begin{centering}
\includegraphics[width=15cm]{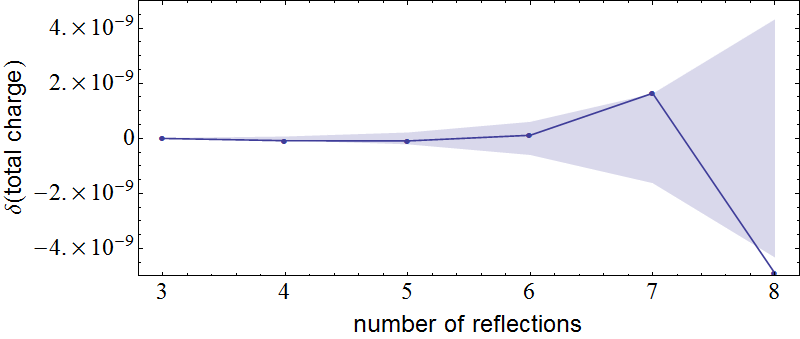}
\par\end{centering}
\caption{The extension of the flat curve from Fig. \ref{4wormsfig1}, extended to $8$ reflections. The grey area shows the estmated numerical error on this quantity (given by $\pm \sqrt{7^n} \times 10^{-16}$, where $10^{-16}$ is approximately the precision used in the calculations, and $7^n$ is the number of mass points in one sphere after $n$ reflection operations). The scale on the $y-$axis has been shifted and rescaled to give a fractional change in the total charge for these larger numbers of reflections.}
\centering{}\label{4wormsfig2}
\end{figure}

In this case, due to the exponential increase in the number of mass points, it is not possible to consider very large numbers of reflections. We have, however, been able to consider what happens to the middle curve in Fig. \ref{4wormsfig1} after $7$ and $8$ reflections. The results of this are displayed in Fig. \ref{4wormsfig2}, together with an estimate of the numerical error involved in the calculation. This error derives from using $\sim 16$ digits of precision in the numerical calculations. It can be seen the numerical results after $8$ reflections are within a few parts in $10^9$ of each other, and that this small change is at the level estimated for the numerical error in the calculation. Despite the difficulty in performing large numbers of reflections, and despite the lack of any obvious convergence in the two of the curves in Fig. \ref{4wormsfig1}, it is therefore still possible to have confidence that we have seen convergence of the total charge inside each reflecting sphere, and that further reflection operations will not significantly change the geometry of the first cosmological region.

%%%%%%%%%%%%%%%%%%%%%%%%%%%%%
%%%%%%%%%%%%%%%%%%%%%%%%%%%%

\subsection{Four Wormholes, and No Flanges}

A space with four wormholes can be created by searching for overlapping coordinate systems in each of the wormhole regions separately. This proceeds just as in the one wormhole case discussed in Sec. \ref{sec:1wh}. Each set of overlapping coordinates within each wormhole is then easily related to each other using the coordinates that span the original cosmological region. It only remains to prove that
\be
\label{JiS2}
J_k \psi_{\infty}^{(4)} = \psi_{\infty}^{(4)} \, ,
\ee
where in this case $\psi_{\infty}^{(4)}$ is the conformal factor that results from a divergent number of reflections in each of our eight spheres, and the index $k$ runs from $1$ to $8$ and labels each of these spheres. The result in Eq. (\ref{JiS2}) follows from the fact that $\psi_{\infty}^{(4)}$ consists of a series of terms that  correspond to every possible series of reflections in every sphere, in the limit that the number of reflections in each of these series diverges. Recalling that $J_k J_k=1$, it can then be seen that adding an extra reflection in every term cannot create any new terms. The LHS of Eq. (\ref{JiS2}) must therefore be equal to $\psi_{\infty}^{(4)}$, as $\psi_{\infty}^{(4)}$ already contains every possible series of reflections.

\section{Discussion}
\label{discussion}

In this paper we have applied the method of images to time-symmetric initial data for vacuum cosmological models. This has allowed us to construct geometries in which multiple cosmological regions are connected together by throats, and has allowed the construction of wormholes that connect antipodal points in a cosmological region. We have applied these methods to the 8-black hole universe \cite{CRT12,BK}, but we could equally well have applied to the other configurations with anti-podal masses, including the 4-black hole universe studied in \cite{Helena}. The scale of the cosmologies that result are similar to those that contain only a single simply connected cosmological region, but allow us to consider the consequences of clustered mass points, and the removal of causally disconnected asymptotically flat regions that would otherwise be required to exist. 

By considering a two-scale problem, we have devised a definition for the interaction energy between clustered mass points in a cosmological model. The interpretation of this energy is not as straightforward as in the asymptotically flat case, as there are no distant regions within the cosmological region that one can go to in order to expand the gravitational field about flat space. Nevertheless, we have been able to use the scale of our cosmological region in order to investigate the gravitational consequences of our interaction energy. We find that when the cluster is small, compared to the scale of the cosmological region, then the total gravitational mass of the cluster appears to be well modelled by the sum total of the proper mass of each of the constituent particles plus the sum of the interaction energies between every pair of particles. If the scale of the cluster is allowed to increase then we find that small inaccuracies are allowed to creep into this simple picture. The consequences of interaction energies between masses that are separated by scales comparable to those of the cosmological region remains to be seen.

We have discussed various different concepts of mass in these models, including the ``proper mass'', the ``effective mass'', and the ``charge'' of each of the objects (see Section 3 for definitions). It has already been shown in previous studies that, if the black holes in such a space-time are sufficiently regularly distributed, that it is the sum of the proper masses that best approximated the total mass in a comparable dust-dominated Friedmann model \cite{CRT12, BK, korzy}. Here we find that this is no longer true when masses are clustered close together (in terms of their angular separation at the centre of the 3-sphere). In this case it is a quantity proportional to the sum of the charges on the poles of the masses that gives the best indication of the contribution of a cluster to the scale of the cosmological model. This quantity also corresponds very closely to the Schwarzschild mass that one would infer from the area of the horizon that encloses the cluster of masses, and provides a way that an observer might operationally try and determine the mass of this system in a given cosmological region. How an observer might try and perform a similar operation when a set of masses are clustered, but not contained within a single enveloping outer horizon remains an open question.

\vspace{1cm}
\noindent {\bf Acknowledgements:} I am grateful to Reza Tavakol and Kjell Rosquist for helpful discussions and comments, and acknowledge support from the STFC.


\section*{References}

\begin{thebibliography}{99}

\bibitem{backreaction}  C.~Clarkson, G.~F.~R.~Ellis, J.~Larena \& O.~Umeh,  {\it Rept. Prog. Phys.} {\bf 74}, 112901 (2011).

\bibitem{backreaction2} T.~Buchert \& S.~R\"{a}s\"{a}nen, {\it Ann. Rev. Nuc. Part. Sci.} {\bf 62}, 57 (2012).

\bibitem{LW} R.~W.~Lindquist \& J.~A.~Wheeler, {\it Rev. Mod. Phys.} {\bf 29}, 432 (1957); {\it erratum} {\bf 31}, 839 (1959).

\bibitem{CF1} T.~Clifton \& P.~G.~Ferreira, {\it Phys. Rev. D} {\bf 80}, 103503 (2009); {\it erratum} {\bf 84}, 109902 (2011).

\bibitem{CF2} T.~Clifton \& P.~G.~Ferreira, {\it JCAP} {\bf 10}, 26 (2009).

\bibitem{CRT12} T.~Clifton, K.~Rosquist \& R.~Tavakol, {\it Phys. Rev. D} {\bf 86}, 043506 (2012).

\bibitem{korzy} M.~Korzy\'{n}ski, {\it Class. Quant. Grav.} {\bf 31}, 085002 (2014).

\bibitem{BK} E.~Bentivegna \& M.~Korzy\'{n}ski, {\it Class. Quant. Grav.} {\bf 29}, 165007 (2012).

\bibitem{CRTG13} T.~Clifton, D.~Gregoris, K.~Rosquist \& R.~Tavakol, {\it JCAP} {\bf 11}, 010 (2013).

\bibitem{CRG14} T.~Clifton, D.~Gregoris \& K.~Rosquist, {\it Class. Quant. Grav.} {\bf 31}, 105012 (2014).

\bibitem{cube1} E.~Bentivegna \& M.~Korzy\'{n}ski, {\it Class. Quant. Grav.} {\bf 30}, 235008 (2013).

\bibitem{cube2} E.~Bentivegna, {\it Class. Quant. Grav.} {\bf 31}, 035004 (2014).
	
\bibitem{cube3} C.-M.~Yoo, H.~Abe, Y.~Takamori \& K.-i.~Nakao, {\it Phys. Rev. D} {\bf 86}, 044027 (2012).

\bibitem{cube4} C.-M.~Yoo, H.~Okawa \& K.-i.~Nakao, {\it Phys. Rev. Lett.} {\bf 111}, 161102 (2013).
	
\bibitem{cube5} C.-M.~Yoo \& H.~Okawa, {\it Phys. Rev. D} {\bf 89}, 123502 (2014).

\bibitem{Misner63} C.~W.~Misner, {\it Annals of Phys.} {\bf 24}, 102 (1963).

\bibitem{gibbons72} G.~W.~Gibbons, {\it Comm. Math. Phys.} {\bf 27}, 87 (1972).

\bibitem{BL63} D.~R.~Brill \& R.~W.~Lindquist, {\it Phys. Rev.} {\bf 131}, 471 (1963).

\bibitem{Kjell} S.~Jolin \& K.~Rosquist, {\it work in progress}.

\bibitem{eisenhart} L.~P.~Eisenhart, ``Riemannian Geometry", Princeton (1925).

\bibitem{Helena} H.~Engstr\"{o}m, ``Properties of Mass Distributions for Discrete Cosmology", Master's Thesis, Stockholm University (2013).






\end{thebibliography}
\end{document}